\documentclass[aps,prb,twocolumn,superscriptaddress,longbibliography]{revtex4-2}

\usepackage[utf8]{inputenc}
\usepackage{physics}
\usepackage{comment}
\usepackage{amssymb}
\usepackage[colorlinks=true,linkcolor=blue,citecolor=blue,urlcolor=blue]{hyperref} 
\usepackage{graphicx}
\usepackage{amsmath}
\usepackage{dcolumn}
\usepackage{bm}
\usepackage{graphicx}
\usepackage{amsmath}
\usepackage{amsfonts}
\usepackage{color}
\usepackage{mathtools}
\usepackage{braket}
\usepackage{gensymb}
\usepackage{textcomp}
\usepackage{soul}

\begin{document}
\title{Anomalous and parallel Hall effects in ferromagnetic Weyl semimetal Cr$_3$Te$_4$}

\author{Anumita Bose}
\affiliation{Solid State and Structural Chemistry Unit, Indian Institute of Science, Bangalore 560012, India.}

\author{Shubham Purwar}
\affiliation{Department of Condensed Matter and Materials Physics, S. N. Bose National Centre for Basic Sciences, Kolkata, West Bengal 700106, India}

\author{Setti Thirupathaiah}
\affiliation{Department of Condensed Matter and Materials Physics, S. N. Bose National Centre for Basic Sciences, Kolkata, West Bengal 700106, India}

\author{Awadhesh Narayan}
\affiliation{Solid State and Structural Chemistry Unit, Indian Institute of Science, Bangalore 560012, India.}

\date{\today}

\begin{abstract}
Recently, time-reversal symmetry broken magnetic Weyl semimetals (WSMs) have attracted extensive attention and have provided an intriguing platform for exploring fundamental physical phenomena. The study of chromium telluride-based systems has also drawn significant interest towards spintronics applications owing to their high Curie temperatures. Here, using \textit{ab initio} calculations, we propose the emergence of multiple Weyl points (WPs) near the Fermi level in such an intrinsic ferromagnetic system, Cr$_3$Te$_4$. The large, well-separated, nontrivial Fermi arcs and surface states, suggest that the WPs are highly robust and resilient to perturbations. A substantial Berry curvature contribution in the vicinity of the Fermi energy not only serves as the origin of large conventional anomalous Hall conductivity (AHC), but also produces unconventional parallel AHC in this material, owing to the low structural symmetry. In addition to the charge Hall conductivity, we also find significant anomalous Nernst conductivities originating from the Berry curvature. Alongside our theoretical predictions, we present complementary experimental results, including X-ray diffraction (XRD) analysis and an examination of the magnetic properties, which demonstrate a Curie temperature of 327 K. Our study advances the understanding of magnetic WSMs, and also encourages further studies in the context of topological properties of our proposed material.
\end{abstract}

\maketitle

\section{Introduction}

The last decade has witnessed tremendous progress in the discoveries of topological phases of matter. After the discovery of topological insulators~\cite{hasan2010colloquium,qi2011topological, moore2010birth}, topological band theory was successfully applied in predicting different topological semimetallic phases ~\cite{yan2017topological,armitage2018weyl} -- Dirac semimetal, Weyl semimetal (WSM) and nodal line semimetals ~\cite{fang2016topological,yu2017topological}.
WSM is a class of topological semimetal, where, valence and conduction bands cross each other, creating two-fold degenerate points near the Fermi level. The band dispersion from the these gapless points is linear in nature and follow the Weyl equation.
In WSMs, these gapless points are known as Weyl points (WPs) and always appear in pairs. These two WPs with opposite chirality act as source (positive chirality) or sink (negative chirality) of Berry flux. The presence of net Berry flux between two WPs in WSM thus causes non closed Fermi arcs connecting the WPs~\cite{wan2011topological}. These arcs can be observed by using surface detecting techniques like angle-resolved photoemission spectroscopy (ARPES)
and scanning tunneling microscopy (STM) -- thus play one of the most important signatures in WSMs. This may lead to a series of exotic properties, such as the chiral anomaly effect~\cite{huang2015observation}, high-mobility carriers~\cite{huang2015observation,takiguchi2020quantum,kaneta2022high}, giant anisotropic optical response, and large negative magnetoresistance ~\cite{arnold2016negative,zhang2016signatures}. At the gapless WPs, the Berry curvature diverges. This large values of berry curvature in turn gives rise to exciting phenomena like giant intrinsic anomalous Hall effects, and anomalous Nernst effect in WSMs.\\
In solid-state systems, the existence of Weyl points requires the breaking of either time-reversal symmetry (TRS) or the inversion symmetry (IS)~\cite{soluyanov2015type,sun2015prediction,huang2016spectroscopic,deng2016experimental,jiang2017signature} or both ~\cite{chang2018magnetic}. Although a myriad of systems with Weyl points due to broken inversion symmetry have been proposed and experimentally realised, systems with broken TRS have been a scarce. External perturbations, like tailoring magnetism in nodal line semimetals~\cite{zhang2018magnetization,nie2020magnetic} and applying hydrostatic pressure~\cite{bose2024pressure} in a trivial magnetic insulator, can give rise to TRS broken WPs. However, finding systems with largely separated WPs and hence with long Fermi arcs has been challenging for a long time. In recent times, TRS broken WPs have been proposed intrinsically in systems such as MnBi$_2$Te$_4$~\cite{zhang2019topological,li2019intrinsic}, Co$_3$Sn$_2$S$_2$~\cite{liu2019magnetic}, EuCd$_2$As$_2$~\cite{wang2019single,soh2019ideal,ma2019spin}, K$_2$Mn$_3$(AsO$_4$)$_3$~\cite{nie2022magnetic}, Ti$_2$MnAl~\cite{shi2018prediction} and BaCrSe$_2$~\cite{sun2023magnetic}.

In recent years, a series of chromium telluride based self-intercalated systems -- Cr$_5$Te$_8$, Cr$_2$Te$_3$, Cr$_3$Te$_4$, and CrTe -- have drawn great attention in the context of two-dimensional magnetism towards potential spintronics applications due to their relatively high T$_c$ and simple binary composition~\cite{zhang2020critical,wang2022layer,wang2023field,purwar2023investigation,matsuoka2024band,goswami2024critical,purwar2024intricate,purwar2024sn0}. However, the study of topological properties of these systems has been lacking so far. Motivated by the recent research progress on Cr-Te-based systems and considering their potential importance for applications, in this work, we study the Cr$_3$Te$_4$ system using \textit{ab-initio} calculation as well symmetry analysis.

We present the occurrence of WPs using first-principles calculations, in  Cr$_3$Te$_4$, with intrinsic ferromagnetic ordering along the (100) direction. Remarkably, the WPs of Cr$_3$Te$_4$ are well-separated, with the Fermi arcs covering as much as 75\% of the reciprocal space vector, hinting towards the robustness of the WPs that are difficult to annihilate by perturbations. We show how symmetry of the crystal becomes a decisive factor in controlling the Berry curvature and hence Hall conductivity components. The large Berry curvature contribution in the vicinity of Fermi energy not only gives rise to large conventional AHC, but also produces unconventional parallel planar AHC, owing to low symmetries in the system. Furthermore, the large Berry curvature also gives rise to anomalous Nernst conductivity (ANC) in the system. We complement our theoretical calculations with experimental synthesis, X-ray diffraction analysis, and magnetic measurements. We find that the T$_c$ for Cr$_3$Te$_4$ is 327 K, which suggests possible room temperature exploration of its topological and magnetic properties.

\begin{figure}
  \includegraphics[scale=0.5250]{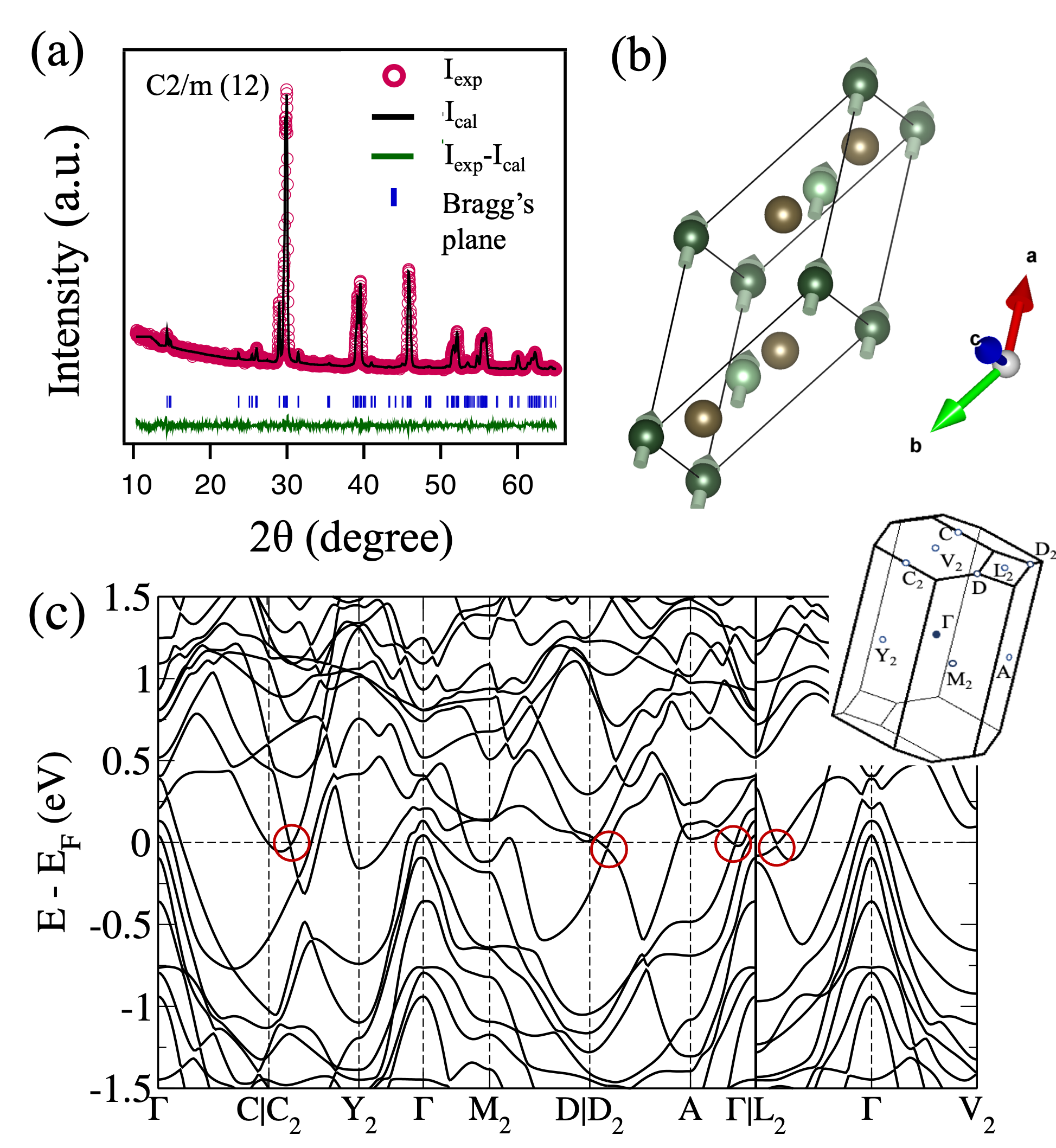}
  \caption{\textbf{Crystal structure and electronic band structure.} (a) X-ray diffraction pattern obtained from crushed single crystals of Cr$_3$Te$_4$, overlaid with the Rietveld refinement results, confirming the monoclinic crystal structure with the space group of C2/m (No. 12). (b) The Primitive unit cell of monoclinic Cr$_3$Te$_4$ consisting of 3 Cr atoms and four Te atoms. Maroon, dark green and light green spheres represent Te, Cr$_I$ and Cr$_{II}$ atoms, respectively. In the ferromagnetic phase, Cr moments are oriented along the $a$ direction. (c) Bulk bandstructure for ferromagnetic Cr$_3$Te$_4$ with SOI, along the high symmetry directions. Nearly degenerate band touching points, which are very close to $E_F$, can be observed along the high symmetry lines C$_2$Y$_2$, D$_2$A, A$\Gamma$ and L$_2\Gamma$, encircled by red circles. We note, however, that the actual band crossings occur slightly away from this high-symmetric path. Inset shows the first BZ for the system, with high symmetry points marked as blue circles.}\label{bulk}
\end{figure}

\section{Methods}

\subsection{Computational methods}

We performed the electronic structure calculations using the
{\sc quantum espresso} package~\cite{giannozzi2009quantum,giannozzi2017advanced}. Generalized gradient approximation is employed based on the Perdew-Burke-Ernzerhof parametrization~\cite{perdew1996generalized}, within projector augmented
wave basis~\cite{blochl1994projector}. Spin-orbit interaction (SOI) is included in the scheme of fully relativistic pseudopotentials. Cutoff values of 50 Ry and 300 Ry were chosen for wave
function and charge density expansions, respectively. For the self-consistent calculation, a $7\times 7\times 7$ Monkhorst-Pack grid~\cite{monkhorst1976special} was used. We used the experimental structure~\cite{Babot1973} to perform the calculations. After the self-consistent calculations, the system was downfolded to an effective tight-binding model with Cr $d$ and Te $p$ as the basis by constructing maximally localized Wannier functions, using the {\sc wannier90} package~\cite{mostofi2008wannier90}. Berry curvature is computed both by Wannier Berri~\cite{tsirkin2021high} and WannierTools~\cite{wu2018wanniertools} codes; the former is for plotting in the three-dimensional Brillouin zone (BZ), while the latter is for plotting in a two-dimensional BZ. Variation of AHC as a function of temperature was calculated by introducing smearing in the WannierTools code~\cite{wu2018wanniertools}. Further ANC was calculated from AHC using Equation~\ref{anc}.

\subsection{Experimental methods}
Cr$_3$Te$_4$ single crystals were synthesized via chemical vapor transport with iodine as the transport agent. A molar ratio of 3.5:4 of chromium (Cr) to tellurium (Te) powders was
used. Thermal treatment occurred over three weeks in a two-zone furnace, with source and
growth zones maintained at 1000$^{\circ}$C and 820$^{\circ}$C, respectively. Powder X-ray diffraction (XRD) analysis utilized a Rigaku X-ray diffractometer (SmartLab, 9 kW) with Cu K$\alpha$ radiation ($\lambda$ = 1.54059 \AA{}). Magnetic measurements ($M(T)$ and $M(H)$) were carried out on a 9 Tesla-PPMS (DynaCool, QuantumDesign).

\section{Results and Discussions}

\subsection{Crystal structure and magnetic measurements}

We begin with our experimental findings. Fig.~\ref{bulk}(a) illustrates the XRD pattern of the crushed Cr$_3$Te$_4$ single crystals, confirming its crystallization in the monoclinic structure with the space group of C2/m (No. 12)~\cite{babot1973proprietes}. The system contains two different types of Cr atoms -- Cr$_I$ and Cr$_{II}$. Cr$_{II}$ atoms form CrTe$_2$ layers and Cr$_I$ atoms take a position in between two such layers (see conventional unit cell in Appendix~\ref{conv}).

For our calculations, we have considered the primitive unit cell of Cr$_3$Te$_4$ as presented in Fig.\,\ref{bulk}(b). The cell parameters are given by $a = b = 7.26$ \AA{} and $c = 6.87$ \AA{}. The comparison of the cell parameters between the experimental data and conventional unit cell is listed in Appendix~\ref{conv}. The unit cell consists of three Cr atoms (one Cr$_I$ type and two Cr$_{II}$ type) and four Te atoms. Cr$_I$, Cr$_{II}$ and Te atoms are represented as dark green, light green and maroon spheres, respectively. In our system ferromagnetically aligned magnetic moments are predominantly present on the Cr atoms are aligned along the $x$ (lattice parameter $a$) direction, which have been represented as light green arrows. The average magnetic moment per Cr$_I$ and Cr$_{II}$ atoms were calculated to be 3.23 and 3.15 $\mu_B$, respectively. 

\begin{figure}
  \includegraphics[scale=0.37]{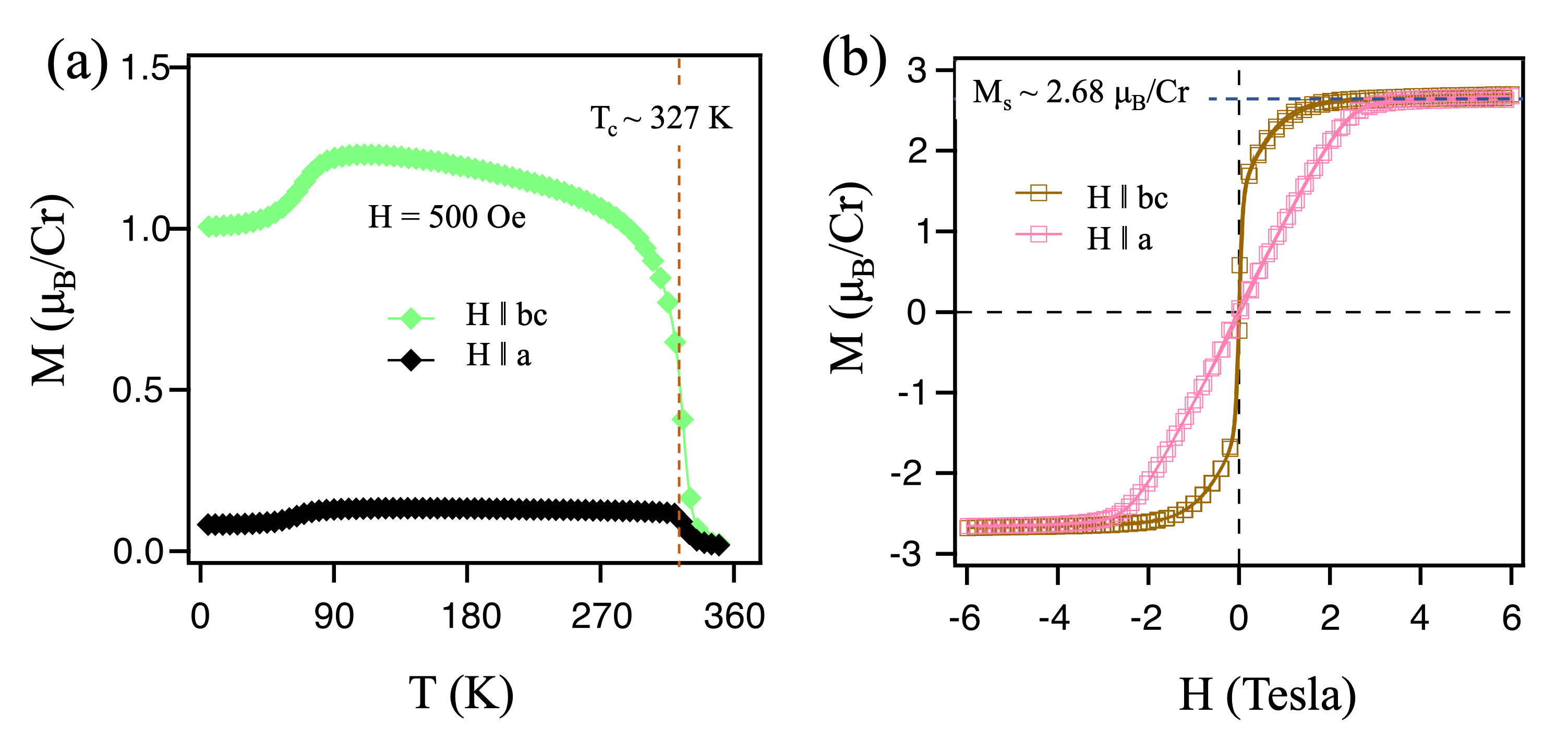}
  \caption{\textbf{Magnetic measurements.} (a) Temperature-dependence of magnetization, $M(T)$, measured under field-cooled (FC) mode with a magnetic field of $H = 500$ Oe for $H \parallel a$ (black line) and $H \parallel bc$ (green line), indicating two magnetic phase transitions in the system. The high temperature transition at 327 K indicates paramagnetic to ferromagnetic transition, while the low temperature transition below 90K seems to have some antiferromagnetic component. (b) Field-dependent magnetization, $M(H)$, measured at $T = 3$ K for magnetic field orientations $H \parallel a$ and $H \parallel bc$, respectively. The saturation magnetization of 2.68 $\mu_B$ is obtained for the Cr atoms.}  \label{expt}
\end{figure}

In Fig.~\ref{expt}, we present the field cooled magnetization variation as a function of temperature, $T$, (in panel (a)) and applied fields, $H$, both for $H \parallel a$ and $H \parallel bc$, as obtained from our experiments. The temperature-dependent magnetization curves in Fig.~\ref{expt} (a) show the presence of two phase transitions -- one at $\sim$ 327 K, representing the ferromagnetic to paramagnetic transition and the other one at 70 K, possibly indicating towards a transition from FM to some antiferromagnetic component below that temperature, consistent with earlier findings~\cite{yamaguchi1972magnetic,zhang2020critical,wang2022layer,goswami2024critical}. Neutron diffraction studies conducted on Cr$_3$Te$_4$ reveal a canted spin configuration where ferromagnetic components tilt relative to the $c$-axis, while antiferromagnetic components align along the ($a + b$)-direction~\cite{andresen1970magnetic}. The interplay between these ferromagnetic and antiferromagnetic phases critically influences the occurrence of multiple magnetic transitions. The field-dependent magnetization measurements performed at 3 K, shown in Fig.~\ref{expt} (b), give a saturation magnetization of 2.68 $\mu_B$, which is slightly lower than the theoretically obtained result.

This centrosymmetric unit cell obeys mirror symmetry $M_y$ in the system, but the orientation of Cr moments along (100) direction breaks this symmetry. However, the system is still invariant under the product between TRS, $\tau$, and the mirror reflection symmetry,  $M_y$, (i.e. $\tau M_y$), even upon inclusion of SOI. Additionally, this magnetic space group also respects the product of $\tau$ and two-fold rotational symmetry about $y$, $\tau C_{2y}$.

\subsection{Bulk electronic structure and Weyl points} 

In Fig.\,\ref{bulk}(c), we present the overview of the bulk electronic band structure for our system, including SOI. The high symmetry points are shown in the inset of Fig.\,\ref{bulk} (c). Presence of the magnetic moments breaks the TRS in the system, which causes lifting of the spin degeneracy and thus leading to band degeneracies at many $k$-points near the Fermi level. These multiple crossings of valence and conduction bands at the $E_F$ makes the system metallic. From the bulk band structure, at a first glance, one can observe seemingly two-fold band degenerate points near $E_F$ along high symmetry directions C$_2$Y$_2$, D$_2$A, A$\Gamma$ and L$_2\Gamma$, which we have highlighted with red circles. Although, upon closer inspection, we find that they reside slightly away from these high symmetry directions. We find that these band degenerate points are indeed Weyl points as they have a finite chirality $\pm 1$. The position of the partner Weyl point in momentum space, with opposite chirality, is controlled by the symmetry operation $\tau M_y$. Under this operation, a point ($\pm k_x$, $\pm k_y$, $\pm k_z$), in the three-dimensional momentum space, transforms as ($\mp k_x$, $\pm k_y$, $\mp k_z$); and chirality, $\chi \longrightarrow -\chi$. These transformations help us to understand that due to the presence of $\tau M_y$, eight non-equivalent Weyl points with chirality $\chi$ at ($\pm k_x$, $\pm k_y$, $\pm k_z$) have their partners at ($\mp k_x$, $\pm k_y$, $\mp k_z$) with chirality $-\chi$. We find that although most of the Weyl points are located very close to each other, two pairs of Weyl points are well separated in momentum space, that can give rise to robustness in surface properties, which we delve into next.

\begin{figure}
\centering
  \includegraphics[scale=0.35]{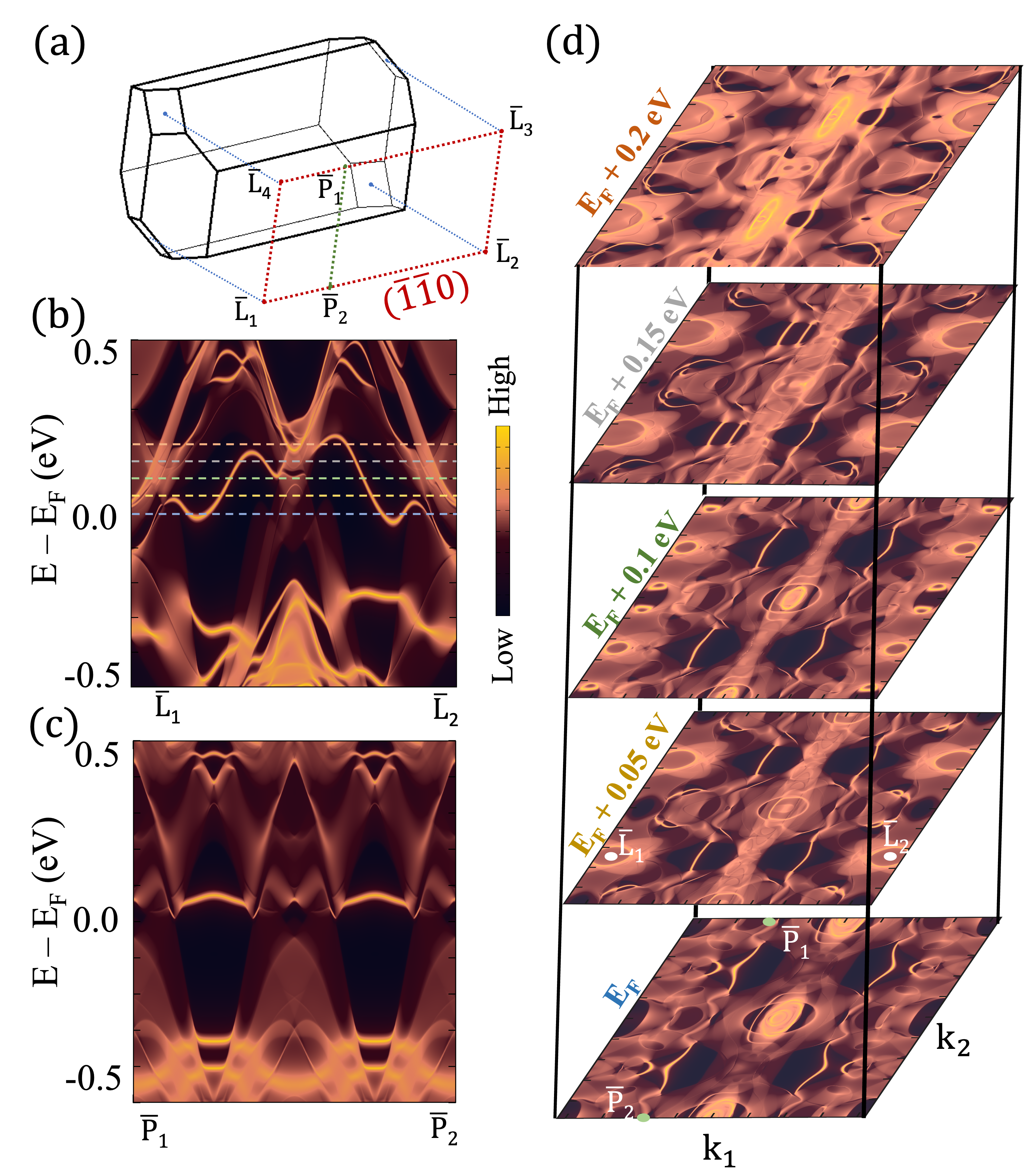}
  \caption{\textbf{Surface topology on ($\Bar{1}\Bar{1}0$) plane.} (a) ($\Bar{1}\Bar{1}0$) surface BZ (marked as red dotted line) projected from the bulk BZ, with high symmetry points $\bar{L}_{1}$, $\bar{L}_{2}$, $\bar{L}_{3}$ and $\bar{L}_{4}$ highlighted. Another non-high symmetric path $\bar{P}_{1}\bar{P}_{2}$ is denoted as the green dotted line. Surface spectra along $\bar{L}_{1}\bar{L}_{2}$ (in (a)) and $\bar{P}_{1}\bar{P}_{2}$ (in (c)) directions. The presence of extended surface states can be observed in (b) and (c). (d) Stacking of constant energy contours with surface state as a function of in-plane momenta. The constant energy surfaces are associated with long, distinct, non-trivial surface arcs confirming the presence of Weyl points, well separated from each other in momentum space.}  \label{fermi_arc}
\end{figure}

\begin{figure*}
  \includegraphics[scale=0.52]{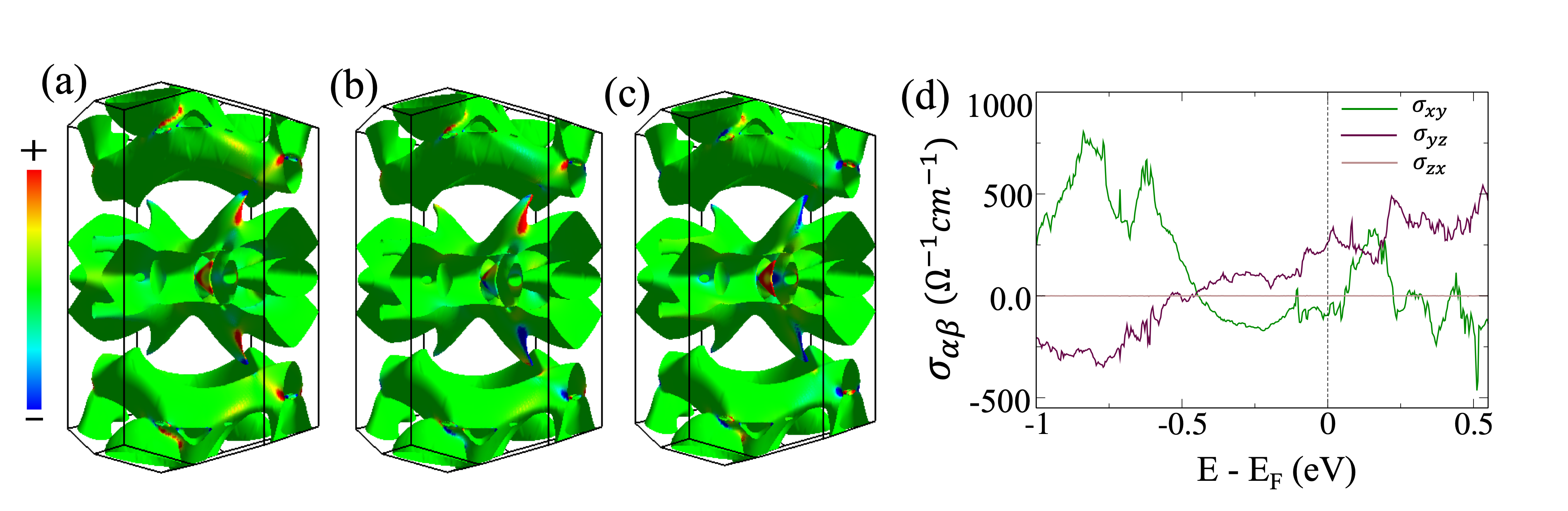}
  \caption{\textbf{Berry curvature and anomalous Hall conductivity.} Berry curvature components (a) $\Omega^x$, (b) $\Omega^y$, and (c) $\Omega^z$ distributed over the Fermi surface over the entire BZ. (d) Variation of AHC components, $\sigma_{\alpha \beta}$, as a function of energy. Green, maroon and light brown curves correspond to $\sigma_{xy}$, $\sigma_{yz}$, and $\sigma_{zx}$, respectively. We note a vanishingly small $\sigma_{zx}$ throughout the energy window. On the other hand, a finite and large value is obtained for $\sigma_{xy}$, $\sigma_{yz}$.}  \label{ahc_bcc}
\end{figure*}

\subsection{Surface analysis and Fermi arcs}

The existence of non-trivial Fermi arcs, i.e., states connecting WPs with opposite chirality, is considered to be one of the most important features of Weyl systems. In order to investigate this property, here, we consider the ($\Bar{1}\Bar{1}0 $) plane for our Cr$_3$Te$_4$ system. Fig.~\ref{fermi_arc}(a) present the surface BZ projected from the bulk BZ. The red dotted line represent the ($\Bar{1}\Bar{1}0 $) surface BZ with the high symmetry points $\bar{L}_{1}$, $\bar{L}_{2}$, $\bar{L}_{3}$ and $\bar{L}_{4}$ marked on it. In Fig.~\ref{fermi_arc}(a) and (b) we show the surface bands along $\bar{L}_{1}\bar{L}_{2}$ and $\bar{P}_{1}\bar{P}_{2}$ directions, respectively. Both of these confirm the presence of extended surface states connecting the bulk bands. Distinct surface states extending from -0.05 eV to 0.23 eV can be observed in Fig.~\ref{fermi_arc}(b). Furthermore, we plot surface states as a function of in-plane momenta for different energy values. These different values of energy are depicted with different colours in Fig.~\ref{fermi_arc}(b). Fig.~\ref{fermi_arc}(d) presents a stack of constant energy contours for five different energy values. As expected from the surface bands, we can clearly observe the presence of long Fermi arcs connecting bulk states. Although a part of the long Fermi arcs merge into the bulk states, a major portion of them is clear, distinct, and well separated from the bulk. Importantly, we also note that the arcs extend to almost the entire $k_2$ vector. Thus, we expect the Fermi arcs in our Cr$_3$Te$_4$ system to be clearly captured in ARPES measurements. Arcs can be also be observed in the wider energy range -0.05 eV to 0.23 eV, although they are prominent in the energy window of 0.0 eV to 0.15 eV. 

In Table.~\ref{arc_comparison}, we present a comparison of arc lengths in previously studied magnetic Weyl materials with our proposed material. We summarize the magnetic ordering, nontrivial surface information along with their energetic position and approximate length of the largest Fermi arcs, as reported in earlier studies. We find that our system hosts arcs having length much larger than that of the non-collinear AFMs Mn$_3$X (X=Ge, Sn) and comparable to to that of Ti$_2$MnAl. The arc separated from the bulk measures 0.46 \AA{}$^{-1}$, which is also sufficiently large. The long, extended nature suggests that these arcs can be detected in future ARPES experiments.

\begin{centering}
\begin{table*}
\caption{\textbf{Comparison of Fermi arc lengths.} Length of Fermi arcs connecting magnetic Weyl points near $E_F$ for different materials along with the magnetic ordering, constant energy and surface hosting the arcs are listed. We find that our system hosts large Fermi arcs on the ($\Bar{1}\Bar{1}0$) surface, comparable to that of Reference~\cite{liu2019magnetic} near $E_F$. $^{**}$ denotes a different kind of surface notation as used by the authors in Ref.~\cite{nie2022magnetic}. The blank places, denoted by `---', indicate missing information about the material.}
\begin{tabular}{p{3.3cm} p{3.4cm} p{2.5cm} p{3.5cm} p{2.5cm}p{2cm}}
\hline
\hline
 System  &Magnetic ordering &Surface & Constant energy for \newline arc, $E_{arc}$ (meV) & Arc length (\AA{}$^{-1}$) & Reference\\
 \hline
 \hline
   Cr$_3$Te$_4$  &FM & $\Bar{1}\Bar{1}0$ & $E_F$& 0.83 & this work \\
 
  Mn$_3$Ge &Non-collinear AFM  &001 & $E_F$+55 & 0.11 & ~\cite{yang2017topological}\\
  
  Mn$_3$Sn &Non-collinear AFM &  001 & $E_F$+86 & 0.10 & ~\cite{yang2017topological}\\
  
  Co$_3$Sn$_2$S$_2$  &FM  &001 & --- &  0.68 & ~\cite{liu2019magnetic}\\
  
  MnBi$_2$Te$_4$ &FM  & $1\Bar{1}0$ & $E_F$+30 &  0.31 & ~\cite{li2019intrinsic}\\
  
  EuCd$_2$As$_2$   &FM&  110 & $E_F$+48 &0.09 & ~\cite{wang2019single}\\
  
  BaCr$_2$Se$_2$  &FM&  001 & --- & 0.33 & ~\cite{sun2023magnetic} \\
  
  K$_2$Mn$_3$(AsO$_4$)$_3$ &FM & $yoz^{**}$ & $E_F$ &  0.44 & ~\cite{nie2022magnetic}\\
  
  Co$_2$ZrSn &FM &001 & $E_F$+500 &  0.25 & ~\cite{wang2016time}\\
  
  Ti$_2$MnAl &Collinear AFM &001 &$E_F$+14 &0.81 & ~\cite{shi2018prediction}\\
\hline
 \hline
 \label{arc_comparison}
\end{tabular}
\end{table*}
\end{centering}

\begin{figure*}
  \includegraphics[scale=0.45]{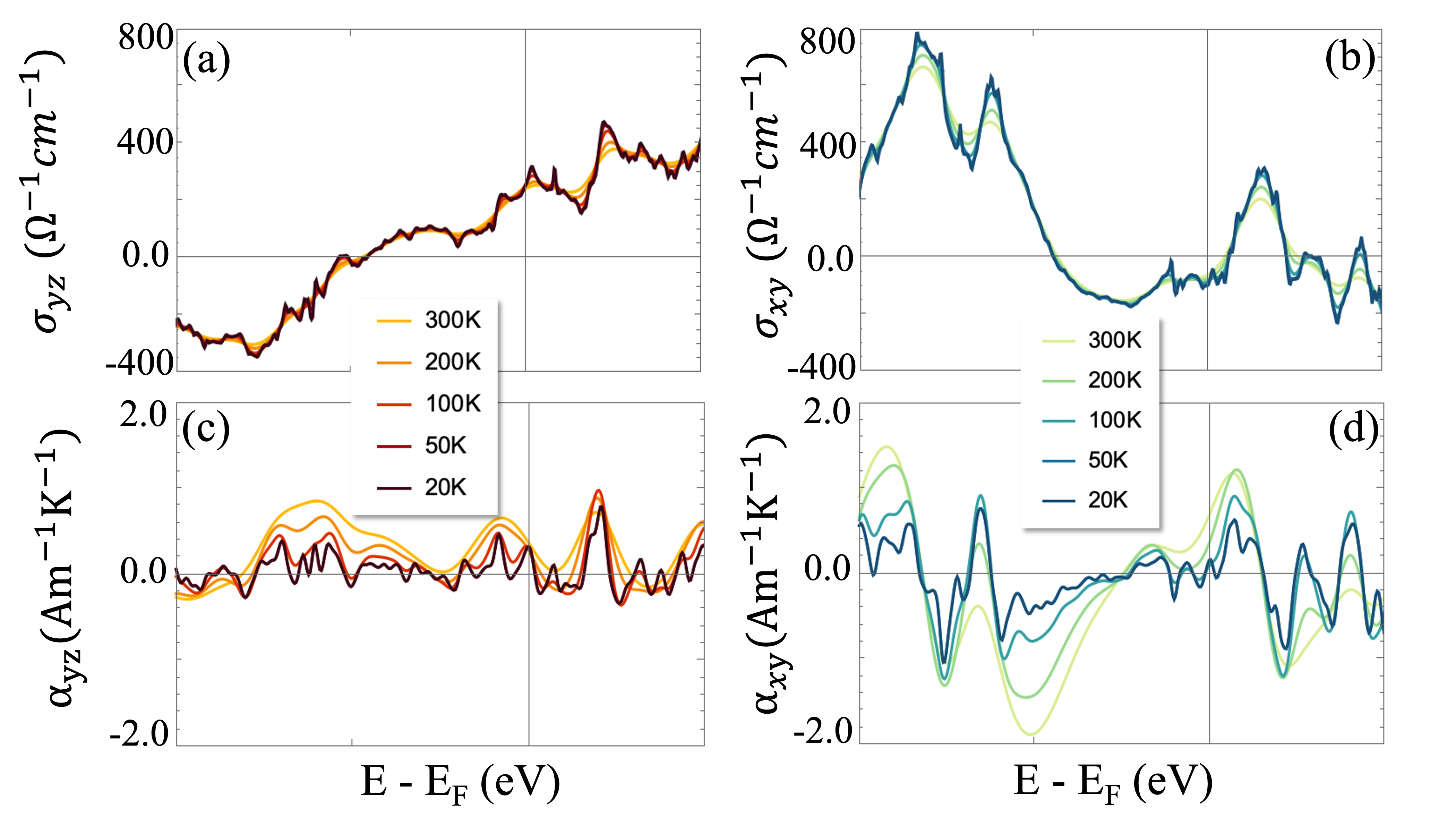}
  \caption{\textbf{Anomalous Hall and anomalous Nernst conductivity.} Top panels show the energy variation for the components of AHC at different values of temperature ranging from 20 K to 300 K -- (a) conventional AHC, $\sigma_{yz}$ and (b) unconventional PAHC, $\sigma_{xy}$. Here zero energy corresponds to the Fermi energy value. Lower panels present the ANC at different temperature values -- (c) $\alpha_{yz}$ and (d) $\alpha_{xy}$ obtained from $\sigma_{yz}$ and $\sigma_{xy}$, respectively. Both the components show a fluctuating nature with the change in the energy. This suggests the tunability of ANC via suitable doping.}  \label{ahc_anc_temp}
\end{figure*}

\subsection{Anomalous Hall effect and the role of symmetries}

Next, we turn our attention to another fascinating property of our system, i.e., large intrinsic AHC. The AHC can be calculated within linear response theory by integrating the Berry curvature, $\Omega$, over the BZ. The $\gamma\delta$-th ($\gamma \neq \delta$) component of the Hall conductivity tensor, $\sigma$, can be written in terms of the $\nu$-th component of $\Omega$ as

\begin{equation}
 \sigma_{\gamma\delta} = \varepsilon_{\gamma\delta\nu}\frac{e^2}{\hbar} \sum_{n} \int_{BZ}^{} f(\epsilon_{n}(\textbf{k})) \Omega_{n}^{\nu}(\textbf{k}) \frac{d\textbf{k}}{(2\pi)^3}.
 \label{sigma}
\end{equation}

Here $e$, $\hbar$, $n$, $\varepsilon$ and $f(\epsilon_{n}(\textbf{k}))$ represent the electron charge, reduced Planck's constant, Bloch band index, Levi-Civita symbol and the Fermi-Dirac distribution function, respectively. Before going into further details of AHC in our sytem, we first examine the nature and components of AHC from a symmetry analysis.

To start with, we consider the symmetry operator $\tau M_y$. Under mirror symmetry $M_y$, the Berry curvature component $\Omega^{y}$ remains unchanged, whereas it is reversed under $\tau$. This enables us to write,

\begin{equation}
\Omega^{y} 
\xrightarrow{\tau M_y} -\Omega^{y}.
\label{omega}
\end{equation}

In other words, $\Omega^{y}$ is odd under $\tau M_y$. This can be seen in Fig.\,\ref{ahc_bcc}(c), where the positive contributions equal the negative ones, resulting in a net zero $\Omega^{y}$ over the BZ. Following Equation~\eqref{sigma}, we obtain $\sigma_{zx} = 0$. To understand the other two transverse Hall conductivity components -- $\sigma_{xy}$ and $\sigma_{yz}$ -- let us focus on the other symmetry present in the system, i.e., $\tau C_{2y}$. Under rotational symmetries, transverse components of $\sigma$ obey the following transformation~\cite{park2022anomalous}

\begin{equation}
\begin{bmatrix} 
\sigma_{yz}  \\
\sigma_{zx}\\
\sigma_{xy} \\
\end{bmatrix}
=  (-1)^p R_m(\theta)  \mathrm{Det}[R_m(\theta)] \begin{bmatrix} 
\sigma_{yz}  \\
\sigma_{zx}\\
\sigma_{xy} \\
\end{bmatrix}.
\end{equation} 

Here $R_{m}(\theta)$ is the three-dimensional matrix for a rotation (both proper and improper) $\theta$ around the axis \textit{\textbf{m}}. The quantity $p$ can take values 0 and 1, which represent the rotation operator associated with and without $\tau$, respectively. For $\tau C_{2y}$, the above equation gives us $\sigma{_{zx}}$ = 0, $\sigma{_{xy}} \neq 0$ and $\sigma{_{yz}} \neq 0$. This is also expected from our calculated Berry curvature components, $\Omega_x$ and $\Omega_z$, as presented in Fig.\,\ref{ahc_bcc}(b) and (d), respectively. Both of these show non-zero contributions over the entire BZ, consistent with the above symmetry analysis.  

Fig.\,\ref{ahc_bcc}(d) shows the components of AHC for our Cr$_3$Te$_4$ system. As anticipated from our symmetry analysis, we find $\sigma_{xz} = 0$ over the entire energy range. We also find non-zero values for the other two transverse components. $\sigma_{yz}$ takes positive values starting from 0.4 eV below $E_F$ to all values above $E_F$, with small variations. At $E_F$, $\sigma_{yz}$ has a value of $\sim$260 $\ohm^{-1}$cm$^{-1}$. Since the magnetic moments of Cr atoms are along the $x$-direction, we expect to obtain a non-zero conductivity in a plane perpendicular to this direction, i.e., $\sigma_{yz}$ is non-zero. This is the conventional AHE, as found in ferromagnets, which is proportional to the magnetization of the system. It is important to note here that the AHC at $E_F$ is larger than the well-studied class of ferromagnetic Co-Fe based Heusler compounds Co$_2$FeX (X = Al, Ga, In, Si, Ge, and Sn), where it takes maximum value of 200 $\ohm^{-1}$cm$^{-1}$ near $E_F$~\cite{huang2015anomalous,noky2018characterization}.

\begin{centering}
\begin{table*}
\caption{\textbf{Comparison of AHC and ANC at $E_F$.} Materials with their magnetic ordering, anomalous Hall conductivities at $E_F$ (both conventional, AHC and unconventional, PAHC) and corresponding Nernst conductivity are listed. The Nernst conductivity for all the systems has been calculated at 300 K. Our proposed material Cr$_3$Te$_4$ hosts AHC value higher than those of the Co-Fe based Heusler compounds. The ANC value for our system is also moderately high, higher than those of the conventional ferromagnets~\cite{weischenberg2013scattering}. The blank places, denoted by `---', indicate missing information about the material.}
\begin{tabular}{p{3.0cm} p{3.2cm} p{2.2cm} p{2.2cm} p{2.0cm} p{2.0cm} p{2cm}}
\hline
\hline
 System &Magnetic ordering &AHC \newline ($\ohm^{-1}$cm$^{-1}$) &PAHC \newline ($\ohm^{-1}$cm$^{-1}$) 
  &ANC \newline (Am$^{-1}$K$^{-1}$) & PANC \newline (Am$^{-1}$K$^{-1}$) & Reference \\
 \hline
 \hline
 
  Cr$_3$Te$_4$ &FM &260 &100& 0.39& 0.72 & this work \\

  Mn$_3$Sn &Collinear AFM &-132& --- &-0.55 & --- &\cite{guo2017large}\\

  Mn$_3$Ge &Collinear AFM &-298&--- &-0.89 &---& \cite{guo2017large}\\

  Mn$_3$Ga &Collinear AFM &-106 &---&2.34-2.41 &---& \cite{guo2017large}\\

   &FM &181 &---&-1.94 &---& \cite{guo2017large}\\

FeCr$_2$Te$_4$& Ferrimagnetic & 130 &50 &---&---&\cite{tan2021unconventional}\\

   Co$_2$FeGe &FM &-78 &--- &3.16&---& \cite{noky2018characterization}\\
   
    Co$_2$FeSn &FM &49 & --- &4.5&---& \cite{noky2018characterization}\\

Fe& FM & 769.8 & --- & 0.17 & --- & \cite{weischenberg2013scattering}\\
Cd& FM & 483.4 & --- & 0.08 & --- & \cite{weischenberg2013scattering}\\
FePt& FM & 833 & --- & 0.62 & --- & \cite{weischenberg2013scattering}\\
FePd& FM & 124.3 & --- & 0.42 & --- & \cite{weischenberg2013scattering}\\
\hline
 \hline
 \label{arc_comparison}
\end{tabular}
\end{table*}
\end{centering}

Interestingly, in addition to the conventional AHC component $\sigma_{yz}$, in Cr$_3$Te$_4$ we find another non-zero component, $\sigma_{xy}$, inspite of having zero net magnetization along the perpendicular direction, i.e., $y$-direction. This phenomenon is known as unconventional parallel anomalous Hall effect (PAHE)~\cite{tan2021unconventional}. This effect, unlike the conventional AHE, appears in the plane of the intrinsic magnetization direction. This phenomenon has been recently reported in experiments in topological semimetal material ZrTe$_5$~\cite{ge2020unconventional}. It is important to mention that the prerequisite to observe PAHE in system is to break rotational or mirror symmetry, since in the presence of such symmetries, Berry curvature can only have components along the magnetization direction, which gives rise only to the conventional AHE~\cite{tan2021unconventional}. The magnetization in the $x$-direction in our system breaks rotational symmetry $C_{2y}$, as well as the mirror symmetry $M_{2y}$, but $\tau C_{2y}$ and $\tau M_{2y}$ remain protected. This allows the system to have a non-zero PAHE in the $xy$-plane. At $E_F$, $\sigma{_{xy}} \sim$100 $\ohm^{-1}$cm$^{-1}$, which is of the same order of magnitude as $\sigma{_{yz}}$. If, by electron doping, the Fermi energy is shifted by 0.14 eV, one can also observe a pronounced peak with a value $\sim$320 $\ohm^{-1}$cm$^{-1}$.

\subsection{Anomalous Nernst effect and Mott relation}

The charge Hall current in a ferromagnet could also be associated with a temperature gradient instead of the usual applied electric field. This is referred to as the anomalous Nernst effect (ANE). ANE is one of the most important mechanisms for exploring the anomalous Hall heat current in ferromagnetic systems. A pioneering study by Xiao \textit{et al.} revealed that such a signal originates from charge carriers with an anomalous transverse velocity in the presence of a longitudinal temperature gradient and a finite Berry curvature~\cite{xiao2006berry}. In this respect, ANE can be considered the thermoelectric counterpart of AHE since both of these originate from the Berry curvature in the momentum space. Once we obtain the chemical potential, $\mu$, dependence of AHC, the ANC components, $\alpha_{\gamma\delta}$, at a temperature $T$, can be calculated using the following relation~\cite{xiao2006berry}

\begin{equation}
\alpha_{\gamma\delta} (\mu) = \frac{1}{e}\int_{-\infty}^{\infty} \sigma_{\gamma\delta}(\epsilon)
\frac{\partial f(\epsilon,\mu)}{\partial \epsilon} \frac{(\epsilon-\mu)}{T} d\epsilon.
\label{anc}
\end{equation}

Our system has a moderately high Curie temperature, $T_{c} \sim 327$ K, which motivated us to calculate ANE for the system. In Fig.\ref{ahc_anc_temp}, we show the variation of AHC as well as ANC as a function of the Fermi energy for different temperature values for an energy range -1 eV to 0.5 eV with respect to the Fermi level. The left panels correspond to the conventional anomalous Nernst conductivity (ANC) and the right panels correspond to the unconventional parallel anomalous Nernst conductivity (PANC). In Fig.~\ref{ahc_anc_temp} (c) and (d), we find that both the ANC components, $\alpha_{yz}$ and $\alpha_{xy}$, show a fluctuating nature as a function of energy. This suggests the tunability of ANC by suitable doping. We will discuss the origin of these features below.

At room temperature, 300 K, $\alpha_{yz} \sim 0.39$ Am$^{-1}$K$^{-1}$. However if the Fermi energy is lowered by 75 meV, $\alpha_{yz}$ becomes $\sim 0.69$ Am$^{-1}$K$^{-1}$. On the other hand $\alpha_{xy}$, in Fig.~\ref{ahc_anc_temp} (d), shows a comparatively higher value at $E_{F}$, which is $\sim 0.72$ Am$^{-1}$K$^{-1}$. This value can further be enhanced to $\sim 1.19$ Am$^{-1}$K$^{-1}$ by a Fermi level shifting of 65 meV via electron doping. It is important to note here that at room temperature ANC values at $E_F$, for our system, are higher than that of conventional ferromagnets Fe ($\sim 0.15$ Am$^{-1}$K$^{-1}$) and Co ($\sim 0.08$ Am$^{-1}$K$^{-1}$), while being comparable to the values for ordered alloys FePt ($\sim 0.6$ Am$^{-1}$K$^{-1}$) and FePd ($\sim 0.42$ Am$^{-1}$K$^{-1}$)~\cite{weischenberg2013scattering}. Although, these values are lower compared to that of Ni ($\sim 1.38$ Am$^{-1}$K$^{-1}$)~\cite{weischenberg2013scattering}. The features of $\alpha_{\gamma\delta}$ as a function of energy can be understood using the Mott relation~\cite{mott,pu2008mott}, as expressed below

\begin{equation}
 \alpha_{\gamma\delta}(\mu) = -\frac{\pi ^2}{3}  \frac{k_{B}^2 T}{e}  \frac{\partial \sigma (\mu)}{\partial \epsilon}E_F,
 \label{mott}
\end{equation}

where $k_{B}$ is the Boltzmann constant. This expression relates $\alpha_{\gamma\delta}$ to the derivative of $\sigma_{\gamma\delta}$ with respect to energy at low temperatures. Therefore, one can expect an enhancement in ANC, whenever there is a change in slope of AHC with energy. In order to understand our results for ANC presented in Fig.~\ref{ahc_anc_temp} (c) and (d), we need to examine the corresponding AHC plots, as presented in Fig.~\ref{ahc_anc_temp} (a) and (b). We find that for low temperature, the peaks in $\alpha_{yz}$ at -75 meV, 190 meV, and -600 meV, with respect to the Fermi level, are associated with the slope changes in $\sigma_{yz}$ at those particular energy values. A similar observation can be made for $\alpha_{xy}$ as well. All the peaks observed for low temperatures e.g., -65 meV, -76 meV, 7 meV, 21 meV and 26 meV are associated with change in curvatures in $\sigma_{xy}$ at these energy values. A careful inspection reveals that the peaks at other energies also have the same origin.\\

\section{Summary} 

To summarize, we have demonstrated a non-trivial Weyl semimetallic phase with multiple Weyl points in ferromagnetically ordered centrosymmetric Cr$_3$Te$_4$, employing first-principles calculations and symmetry arguments. As a signature of these non-trivial bulk Weyl points, we observe large, distinct and robust Fermi arcs on ($\Bar{1}\Bar{1}0$) surface. These arcs occupy almost entire reciprocal lattice vector and more than half are well separated from the bulk states, and are observed in the energy window ranging from 0.0 eV to 0.15 eV. We expect these to be captured in ARPES and STM experiments in the near future. We have also shown that the system exhibits large intrinsic conventional AHC $\sigma_{yz}$, originating from a large Berry curvature. In addition to the conventional out of plane AHE, the system also exhibits an intrinsic PAHE, with value of $\sim$100 $\ohm^{-1}$cm$^{-1}$ at $E_F$, protected by its crystal symmetries. We performed ANC calculations (both conventional and unconventional) for temperature values ranging from 20 K to 300 K. Both $\alpha_{yz}$ and $\alpha_{xy}$ for our system show moderately large values. At 300 K, our calculated values at the Fermi level turn out to be greater than ferromagnetic Fe and Co. In addition to our theoretical predictions, we provide complementary experimental findings for our system. These include XRD study and a study of the magnetic properties, which reveal a Curie temperature of 327 K. Therefore, we expect the exploration of topological and magnetic properties of Cr$_3$Te$_4$ at room temperature in the near future.

\section*{Acknowledgement}
AB thanks Temuujin Bayaraa for useful discussions. AB acknowledges Prime Minister's Research Fellowship for financial support. AN acknowledges support from ANRF (project number CRG/2023/000114).

\appendix
\section{Conventional Unit Cell}
\label{conv}




Cr$_3$Te$_4$ crystallizes in the monoclinic space group C2/m (No. 12). In the main text, we have presented the structural details for the primitive unit cell that have been used for our computations. Here, we show the same for the conventional unit cell for a better visualization and understanding the symmetries. Table~\ref{table:unitcell} displays the comparison between the lattice parameters for the conventional unit cell for the structure used in our first-principles calculation and obtained from our experiments.

\begin{figure}[h!]
  \includegraphics[scale=0.16]{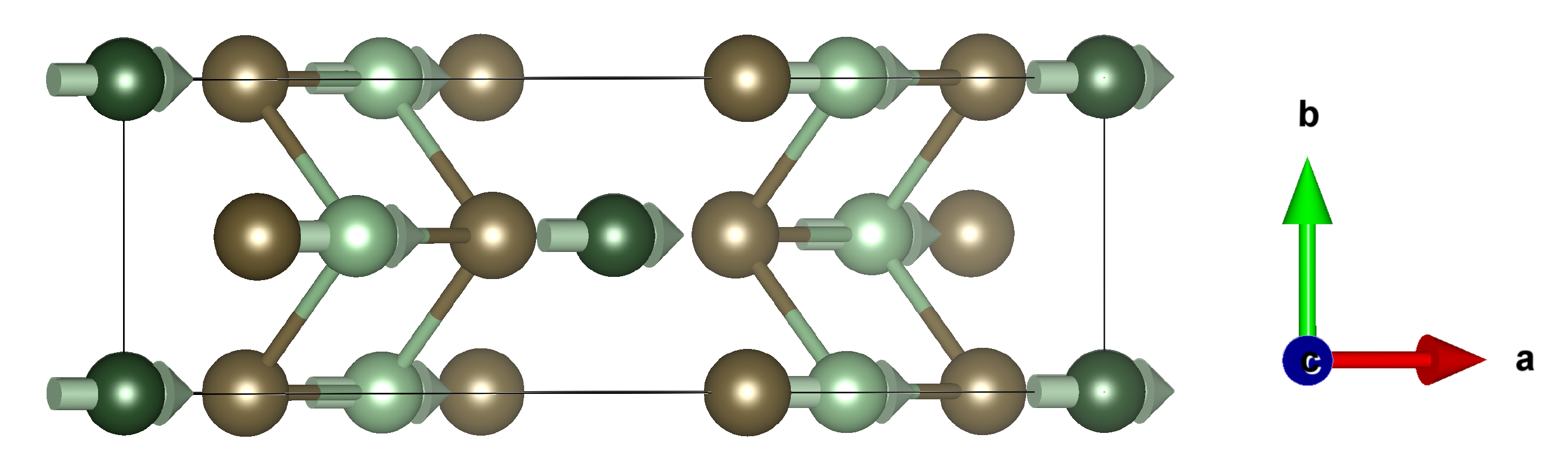}
  \caption{\textbf{Conventional unit cell of Cr$_3$Te$_4$.} Magenta, green and yellow spheres represent Cr$_I$, Cr$_{II}$ and Te atoms, respectively. The crystal structure of Cr$_3$Te$_4$ can be considered as a stacking of CrTe layers (with Cr$_{II}$) along the $a$ direction. Cr$_{I}$ atoms are intercalated between two such CrTe$_2$ layers.}  \label{conventional}
\end{figure}

\begin{table}[h!]
\caption{\textbf{Comparison of cell parameters.} }
\label{table:unitcell}
\begin{tabular}{p{2.5cm} p{1.8cm} p{1.8cm}p{1.8cm}}
\hline
\hline
 Structure & $a$ (\AA{})& $b$ (\AA{}) & $c$ (\AA{})\\
 \hline
 \hline
 Theoretical &13.97 & 3.84 &6.87\\
  Experimental  & 14.02 & 3.94 &  6.87\\
\hline
 \hline
\end{tabular}
\end{table}

\bibliography{references}

\end{document}